\documentclass{ws-mpla}

\usepackage{graphicx}
\usepackage{axodraw}

\newcommand{\imag}{\Im {\rm m}}
\newcommand{\real}{\Re {\rm e}}

\def\gsim{\:\raisebox{-0.5ex}{$\stackrel{\textstyle>}{\sim}$}\:}

\begin{document}

\markboth{Jae Sik Lee}
{Probing Higgs-sector CP Violation at a Photon Collider}

\catchline{}{}{}{}{}

\title{PROBING HIGGS-SECTOR CP VIOLATION  \\ AT A PHOTON COLLIDER }

\author{\footnotesize JAE SIK LEE }

\address{
Center for Theoretical Physics, School of Physics and Astronomy, \\
Seoul National University, Seoul 151-747, Korea\\
jslee@muon.kaist.ac.kr}

\maketitle


\begin{abstract}
In this review we demonstrate physics potential of a photon linear collider
by studying the neutral Higgs-boson sector of the MSSM in which
interesting CP-violating Higgs mixing could arise via radiative corrections.

\end{abstract}

\ccode{PACS Nos.: 14.80.Cp, 12.60.Jv, 11.30.Er}

\section{Introduction}	

The two-photon collision option of the International Linear $e^+e^-$ Collider (ILC), 
a photon linear collider (PLC), is using 
back-scattered laser photons off the incident 
electrons and/or positrons. \cite{GKPST,TESLATDR_Photon}
A photon collider provides unique capabilities for probing
new physics appearing
in neutral spin-zero particles such as neutral Higgs bosons in the Standard
Model (SM) and its Minimal Supersymmetric extension, MSSM. \cite{Hagiwara:2000bt}
The fact that the two-photon energy can reach about 80 \% of 
the $e^+e^-$ center-of-mass (c.m.) energy and
the two-photon luminosity can be comparable to or even larger than
that of $e^+e^-$ collisions
makes it easy to produce Higgs bosons 
copiously as $s$-channels resonances at the PLC.
%
Especially, a capability of controlling polarizations of the colliding photons
makes the PLC an ideal machine to probe CP properties of the Higgs bosons.
In this review, to demonstrate physics potential of the PLC,
we investigate the neutral Higgs-boson sector of the MSSM in which
CP-violating mixing among scalar and pseudo-scalar states
could be radiatively induced.

This review is organized as follows. 
Section \ref{sec:plc}  describes a basic mechanism of
generating polarized back-scattered laser photons at the ILC. We explicitly show 
the helicity-dependent $\gamma\gamma$ luminosity as well as
the mean polarization of two colliding photon beams. 
In Sec.~\ref{sec:mssmcpv}, we briefly review the Higgs-sector CP violation in the MSSM 
induced by the CP-violating phases of the soft SUSY breaking terms.
Section \ref{sec:schannel} and Section \ref{sec:pptoff} are devoted to CP violation 
in $s$-channel production of Higgs bosons and 
in the process $\gamma\gamma \rightarrow f \bar{f}$, respectively.
Section \ref{sec:con} summarizes our conclusions.
 
\section{A Photon Collider}
\label{sec:plc}
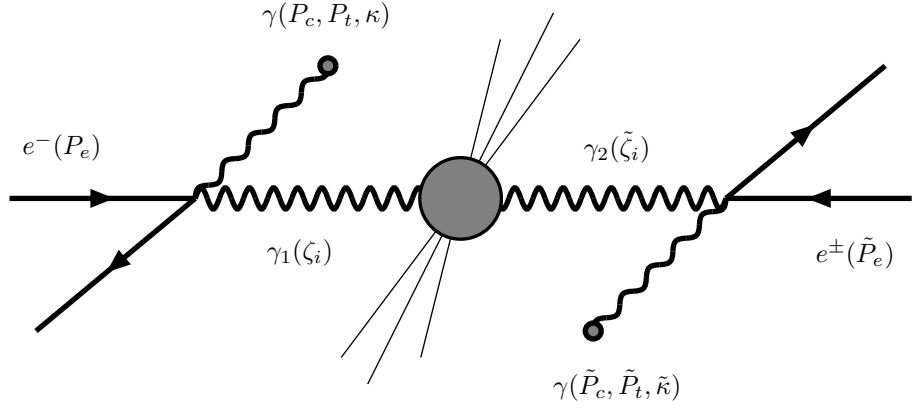
\begin{figure}[htbh]
\vspace*{-3cm}
\begin{center}
\begin{picture}(400,300)(0,0)

\SetWidth{2.0}
\ArrowLine(30,150)(100,150)
\Photon(100,150)(200,150){4}{11}
\Photon(200,150)(300,150){4}{11}
\ArrowLine(370,150)(300,150)
\ArrowLine(100,150)(40,100)
\ArrowLine(300,150)(360,200)
\Photon(100,150)(150,200){2}{5}
\Photon(300,150)(250,100){2}{5}
\GOval(150,200)(3,3)(360){0.5}
\GOval(250,100)(3,3)(360){0.5}
\GOval(200,150)(15,15)(360){0.5}
\SetWidth{0.5}
\Line(200,150)(235,220)
\Line(200,150)(165,80)
\Line(200,150)(245,210)
\Line(200,150)(155,90)
\Line(200,150)(215,210)
\Line(200,150)(185,90)
\GOval(200,150)(15,15)(360){0.5}
\Text(50,170)[]{${e^- (P_e)}$}
\Text(350,130)[]{${e^{\pm} (\tilde{P}_e)}$}
\Text(150,220)[]{${\gamma (P_c,P_t,\kappa)}$}
\Text(260,80)[]{${\gamma (\tilde{P}_c,\tilde{P}_t,\tilde{\kappa})}$}
\Text(140,130)[]{${\gamma_1 (\zeta_i)}$}
\Text(260,170)[]{${\gamma_2 (\tilde{\zeta}_i)}$}
\end{picture}\\
\end{center}
\vspace{-2.5cm}
\caption{{\it  Schematic diagram 
of a $\gamma\gamma$ collider. Polarizations of the initial
electron and positron beams are denoted by $P_e$ and $\tilde{P}_e$, respectively. 
For initial laser photons: $P_c$ and $\tilde{P}_c$ are 
degrees of circular polarization or mean photon helicity, 
$P_t$ and $\tilde{P}_t$ the degrees of linear polarization, and
$\kappa$ and $\tilde{\kappa}$ the azimuthal angles of the direction of 
the maximum linear polarizations.
The mean polarizations of the colliding photon beams are denoted by the
Stokes parameters $\zeta_i$ and $\tilde{\zeta}_i$, 
see Eq.~(\protect\ref{eq:zeta}).
} }
\label{fig:ppcollider}
\end{figure}
\noindent
The ideal luminosity of the PLC, from the Compton back-scattering of
laser photons, is given by \cite{GKPST}
\begin{equation}
\frac{{\rm d}^2{\cal L}_{\gamma\gamma}}{{\rm d}y_1{\rm d}y_2}=
f(y_1)f(y_2) \,{\cal L}_{ee}
\label{eq:gglum0}
\end{equation}
where 
\begin{equation}
y_i\equiv E_{\gamma_i}/E_b
\end{equation}
with the electron beam energy
$E_b= \sqrt{s}/2$. Here $s$ is the invariant ILC
energy squared and $y_i$ vary between $0$ and $y_{\rm max}=x/(1+x)$. The
machine parameter $x$ depends on $E_b$ and the laser-photon energy $\omega_0$:
\begin{equation}
x = \frac{4\, E_b\, \omega_0}{m_e^2}=
15.3\left(\frac{E_b}{\rm TeV}\right)\left(\frac{\omega_0}{\rm eV}\right)\,.
\end{equation}
The function $f(y_1)$ in Eq.~(\ref{eq:gglum0}) is given by
\begin{equation}
f(y_1)=\frac{1}{\sigma_c}\frac{d\sigma_c}{dy_1}
=\frac{2\pi\alpha^2}{\sigma_c x m_e^2}\left[
1/(1-y_1)+(1-y_1)-4r(1-r)-r x (2r-1)(2-y_1) P_eP_c \right]
\label{eq:fy}
\end{equation}
where
\begin{equation}
r\equiv \frac{y_1}{x(1-y_1)}\rightarrow 1 \ \ {\rm as} \ \
y_1\rightarrow y_{\rm max}
\end{equation}
and
$P_e$ and $P_c$ are polarizations of the initial electron and laser photon beams,
respectively, see Fig.~\ref{fig:ppcollider}. 
For example, $(P_e,P_c)=(+1,+1)$ means the right-handed electron and 
laser photon. 
For a second photon $\gamma_2$, $f(y_2)$ can be obtained by replacing
$(P_e,P_c)$  with $(\tilde{P}_e,\tilde{P}_c)$.
The total Compton cross-section $\sigma_c$ in Eq.~(\ref{eq:fy}), 
which is independent of $y$ but depends on 
the initial electron (positron) and laser-photon polarizations, 
is given by
\begin{eqnarray}
\sigma_c &\equiv& \sigma_c^0+P_eP_c\,\sigma_c^1\,,
\nonumber \\
\sigma_c^0 &=& \frac{2\pi\alpha^2}{x m_e^2}
\left[\left(1-\frac{4}{x}-\frac{8}{x^2}\right)\log(x+1)
+\frac{1}{2}+\frac{8}{x}-\frac{1}{2(x+1)^2}\right]\,,
\nonumber \\
\sigma_c^1 &=& \frac{2\pi\alpha^2}{x m_e^2}
\left[\left(1+\frac{2}{x}\right)\log(x+1)
-\frac{5}{2}+\frac{1}{x+1}-\frac{1}{2(x+1)^2}\right]\,.
\label{eq:compton}
\end{eqnarray}
The photon-photon luminosity (\ref{eq:gglum0}) can be rewritten as
\begin{equation}
\frac{1}{{\cal L}_{ee}}\frac{{\rm d}{\cal L}_{\gamma\gamma}}{{\rm d}\tau}=
\int^{y_{\rm max}}_{\tau/y_{\rm max}}\frac{{\rm d}y'}{y'}
f(y')f(\tau/y')
\label{eq:gglum1}
\end{equation}
where $\tau\equiv \hat{s}/s$ with 
$\hat{s}$ being the c.m. energy of colliding photons $\gamma_1$ and $\gamma_2$.
 
\begin{figure}[htbh]
\vspace{-1.0cm}
\centerline{\epsfig{figure=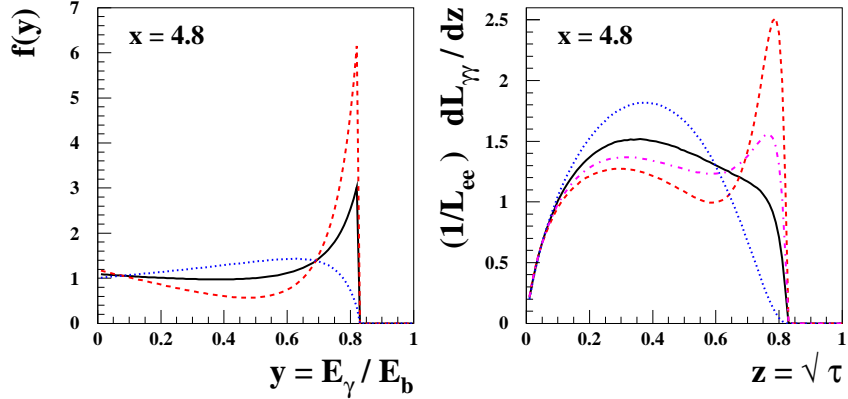,height=12cm,width=12cm}}
\vspace{-5.5cm}
\caption{{\it Photon energy spectrum (left frame) and spectral luminosity 
of $\gamma\gamma$ collisions (right frame).  In the left frame, the solid black line 
is for $P_e\cdot P_c=0$, the dashed red line for $P_e\cdot P_c=-1$, and the dotted blue 
line for $P_e\cdot P_c=1$.  In the right frame, 
$(P_e\cdot P_c,\widetilde{P}_e\cdot\widetilde{P}_c)=(0,0)$, $(-1,-1)$, $(1,1)$, 
and $(0,-1)$ for the solid black, dashed red, dotted blue, and dash-dotted magenta 
lines, respectively.  
}}
\label{fig:fxylum}
\end{figure}
In Fig.~\ref{fig:fxylum}, we show $f(y)$ as a function of $y=E_\gamma/E_b$
and $\frac{1}{{\cal L}_{ee}}\frac{d{\cal L}_{\gamma\gamma}}{dz}$ as a function of
$z\equiv \sqrt{\tau}$ for various combinations of initial polarizations
$P_e\cdot P_c$ and $\tilde{P}_e\cdot \tilde{P}_c$ taking $x=4.8$.
Note that the functions
with $P_e\cdot P_c=\tilde{P}_e\cdot \tilde{P}_c=-1$ (dashed lines)
have peaks in the hard photon region.

The polarization of the colliding photon beam is given by the
Stokes parameters $\zeta_i$ ($\tilde{\zeta}_i$) 
which describe polarization transfer from the initial
laser light and electron (positron)
to the colliding photon: $\zeta_2$ is the degree of circular
polarization and $(\zeta_3,\zeta_1)$ are the degrees of linear polarization
transverse and normal to the plane defined by the electron direction and
the direction of the maximal linear polarization of the initial laser light.
The mean polarization is given by 
\begin{equation}
\zeta_i = \frac{C_i}{C_0}
\label{eq:zeta}
\end{equation}
where, explicitly,
\begin{eqnarray}
C_0&=&\frac{1}{1-y}+1-y-4r(1-r)-rx(2r-1)(2-y)P_eP_c
= \left(\frac{2\pi\alpha^2}{\sigma_c x m_e^2}\right)^{-1} f(y) \nonumber \\
C_1&=&2r^2 P_t \sin 2\kappa\nonumber \\
C_2&=&rx[(1+(1-y)(2r-1)^2]P_e-(2r-1)(\frac{1}{1-y}+1-y)P_c \nonumber \\
C_3&=&2r^2 P_t \cos 2\kappa
\label{eq:Cs}
\end{eqnarray}
where $P_c$, $P_t$, and $\kappa$ are for the polarization states of the laser
photon.  Similar relations can be obtained for
$\tilde{\zeta_i}$ by replacing $(P_c, P_t, \kappa)$ with
$(\tilde{P}_c, \tilde{P_t}, -\tilde{\kappa})$, see Fig.~\ref{fig:ppcollider}.

\begin{figure}[htbh]
\vspace{-1.5cm}
\centerline{\epsfig{figure=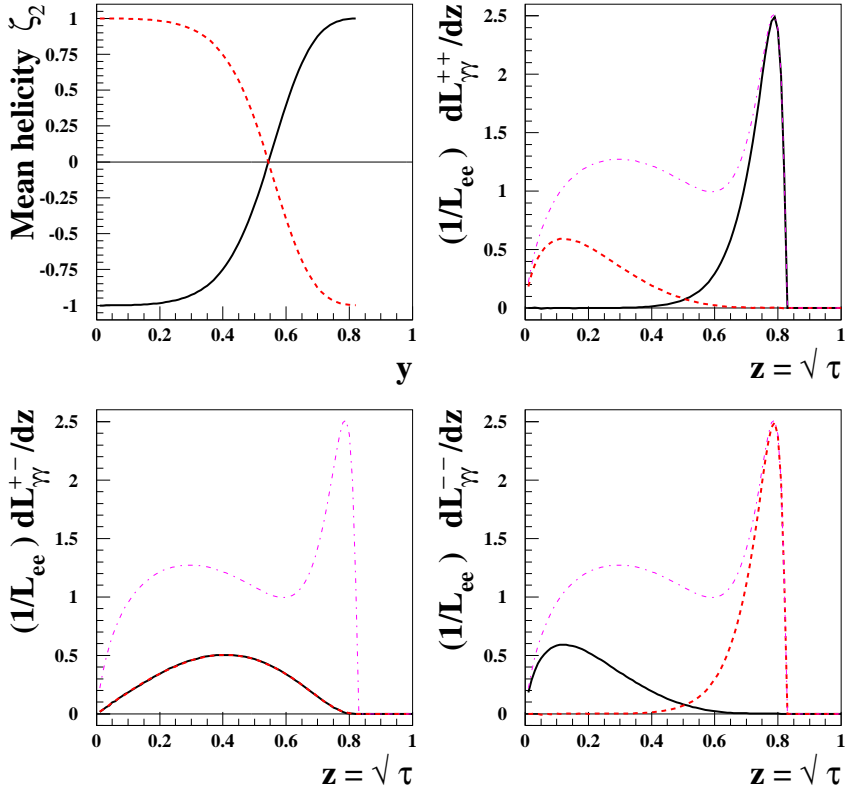,height=12cm,width=12cm}}
\vspace{-0.5cm}
\caption{{\it
The mean helicity of the colliding photon beam $\zeta_2$
and the $\gamma\gamma$ luminosity functions
$\left(1/{\cal L}_{ee}\right){\rm d}{\cal L}^{\lambda_1\lambda_2}/{\rm d}z$
when $P_e\cdot P_c=\widetilde{P}_e\cdot \widetilde{P}_c=-1$ and $x=4.8$.
The upper indices of the luminosity functions denote the helicity combinations
of the colliding photon beams.  The solid (black) and dashed (red)
lines are for $P_e=\widetilde{P}_e=+1$ and $-1$, The thin dash-dotted
(magenta) lines are for the unpolarized luminosity given by
Eq.~(\ref{eq:gglum1}) or 
$(1/{\cal L}_{ee})\sum_{\lambda_1\,,\lambda_2=\pm}{\rm d}{\cal L}^{\lambda_1\lambda_2}/{\rm d}z$.
 }}
\label{fig:pol}
\end{figure}

The mean helicity of the colliding photon beam $\zeta_2 $
and the $\gamma\gamma$ luminosity functions for four combinations of
helicities are shown in Fig.~\ref{fig:pol}
when $P_e\cdot P_c=\widetilde{P}_e\cdot \widetilde{P}_c=-1$ taking $x=4.8$.
The $\gamma\gamma$ luminosity function for the helicity combination
$(\lambda_1,\lambda_2)$ is given by
\begin{equation}
\frac{1}{{\cal L}_{ee}}
\frac{{\rm d}{\cal L}^{\lambda_1\lambda_2}}{{\rm d}\sqrt{\tau}}
=\left(\frac{ \langle 00 \rangle _\tau
+\lambda_1 \langle 20 \rangle_\tau
+\lambda_2 \langle 02 \rangle_\tau
+\lambda_1\lambda_2 \langle 22 \rangle_\tau } {4}\right)
\left(\frac{2\pi\alpha^2}{\sigma_c^{\gamma_1} x m_e^2}\right)
\left(\frac{2\pi\alpha^2}{\sigma_c^{\gamma_2} x m_e^2}\right)
\end{equation}
where $\sigma^{\gamma_i}_c$ is the Compton scattering cross-section given in
Eq.~(\ref{eq:compton}) and the polarization-correlation is defined as
\begin{equation}
\langle ij\rangle_\tau \equiv  2 \sqrt{\tau}
\int^{y_{\rm max}}_{\tau/y_{\rm max}}\frac{{\rm d}y'}{y'} C_i(y')C_j(\tau/y')\,.
\end{equation}
From Fig.~\ref{fig:pol}, we observe 
that the mean helicity of a colliding photon is the
same as that of the electron at high energy in this case, i.e.
$P_e\cdot P_c=\widetilde{P}_e\cdot \widetilde{P}_c=-1$.
Also note that the luminosity functions
for $(\lambda_1\,,\lambda_2)=(\pm\,,\pm)$ have peaks at high $z$
when $P_e=\widetilde{P}_e=\pm$, see the solid (dashed) line in 
the upper-right (lower-right) frame of Fig.~\ref{fig:pol}.

\section{CP Violation in the MSSM Higgs Sector}
\label{sec:mssmcpv}

Any phenomenologically viable SUSY model requires {\it soft} SUSY breaking terms 
which break SUSY without spoiling the SUSY resolution to
the hierarchy problem. There are three types of soft SUSY breaking terms 
appearing in the MSSM:

\begin{itemize}
\item \underline{Gaugino mass terms}:
\begin{equation}
\frac{1}{2}\left(M_3\,\widetilde{g}^a\widetilde{g}^a+M_2\,
\widetilde{W}^i\widetilde{W}^i
+M_1\, \widetilde{B}\widetilde{B} +{\rm h.c.} \right)\,,
\label{eq:gaugino}
\end{equation}
where $M_3$ is a gluino mass parameter of the gauge group SU(3)$_c$
and $M_2$ and $M_1$ are wino and bino mass parameters of the gauge groups
SU(2)$_L$ and U(1)$_Y$, respectively.
\item \underline{Trilinear $A$ terms}:
\begin{equation}
 \widetilde{u}^*_R\,h_u\,A_u\,\widetilde{Q}H_2
-\widetilde{d}^*_R\,h_d\,A_d\,\widetilde{Q}H_1
-\widetilde{e}^*_R\,h_e\,A_e\,\widetilde{L}H_1+{\rm h.c.}\,,
\label{eq:aterms}
\end{equation}
where $\widetilde{Q}$ and $\widetilde{L}$ are SU(2)$_L$ doublet squark and
 slepton
fields and $\widetilde{u}_R$, $\widetilde{d}_R$, and $\widetilde{e}_R$ are
SU(2)$_L$ singlet fields.
\item \underline{Scalar mass terms}:
\begin{eqnarray}
&&
\widetilde{Q}^\dagger \, { M^2_{\widetilde{Q}}}\,\widetilde{Q}
+\widetilde{L}^\dagger \,{ M^2_{\widetilde{L}}}\,\widetilde{L}
+\widetilde{u}_R^* \,{ M^2_{\widetilde{u}}}\, {\widetilde{u}_R}
+\widetilde{d}_R^* \,{ M^2_{\widetilde{d}}}\, {\widetilde{d}_R}
+\widetilde{e}_R^* \,{ M^2_{\widetilde{e}}}\, {\widetilde{e}_R}
\label{eq:scalarmasses}
\nonumber \\
&&
+ \, {m_2^2} H_2^* H_2
+ {m_1^2} H_1^* H_1
-({m_{12}^2} H_1 H_2 + {\rm h.c.})\,. \\ \nonumber
\end{eqnarray}
\end{itemize}
\vspace{-0.5cm}
One crucial observation is that all the massive parameters appearing in the
soft SUSY breaking terms can be complex containing nontrivial
CP-violating phases. Together with the phase of the Higgsino mass
parameter $\mu$ of the term $-\mu H_1 H_2$ in the superpotential,
all the physical observables depend on the CP phases of the combinations
${\rm Arg}[M_i\,\mu\,(m_{12}^2)^*]$ and
${\rm Arg}[A_f\,\mu\,(m_{12}^2)^*]$. \cite{Dugan:1984qf,Dimopoulos:1995kn}
\footnote{In this review, we are going to take the convention of
${\rm Arg}(m_{12}^2)=0$ while keeping the explicit dependence of ${\rm Arg}(\mu)$.}

An interesting phenomenological result of the CP-violating phases 
is that loop effects mediated dominantly by third-generation squarks
with large Yukawa couplings
may lead to sizeable violation of the tree-level CP invariance of the MSSM
Higgs potential, giving rise to significant Higgs
scalar--pseudoscalar transitions.\cite{Pilaftsis:1998pe,Pilaftsis:1998dd,CPNSH}
The size of the CP-violating mixing is proportional to
\begin{equation}
\frac{3}{16\pi^2}\,
\frac{\Im{\rm m}(A_f\mu)}{m^2_{\tilde{f}_2}-m^2_{\tilde{f}_1} }
\label{eq:sizeofmixing}
\end{equation}
with $f=t,b$.
At two-loop level, the gluino mass parameter becomes relevant, for example, through the
possibly important threshold corrections to the top- and bottom-quark Yukawa
couplings.  

One consequence of the CP-violating mixing among three MSSM neutral 
Higgs bosons is that the couplings of Higgs bosons to the SM and SUSY particles 
are significantly modified. Accordingly the loop-induced Higgs boson couplings to 
two photons, which receive contributions from all types of charged particles, are
largely affected.
The amplitude for the production process
$\gamma\gamma\rightarrow H_i$ can be written as
\begin{eqnarray} \label{eq:hipp}
{\cal M}_{\gamma\gamma H_i}=-\frac{\alpha\hat{s}}{4\pi\,v}
\bigg\{S^\gamma_i(\hat{s})\, \left(\epsilon_{1\perp}\cdot\epsilon_{2\perp}\right)
 -P^\gamma_i(\hat{s})\frac{2}{\hat{s}}
\langle\epsilon_1\epsilon_2 k_1k_2\rangle
\bigg\}\,,
\end{eqnarray}
where $k_{1,2}$ are the momenta of the two photons with $(k_1+k_2)^2=\hat{s}$ and
$\epsilon_{1,2}$ the wave vectors of the corresponding photons,
$\epsilon^\mu_{1\perp} = \epsilon^\mu_1 - 2k^\mu_1 (k_2 \cdot
\epsilon_1) / \hat{s}$, $\epsilon^\mu_{2\perp} = \epsilon^\mu_2 -
2k^\mu_2 (k_1 \cdot \epsilon_2) / \hat{s}$ and $\langle \epsilon_1
\epsilon_2 k_1 k_2 \rangle \equiv \epsilon_{\mu\nu\rho\sigma}\,
\epsilon_1^\mu \epsilon_2^\nu k_1^\rho k_2^\sigma$. The scalar and
pseudoscalar form factors, retaining only the dominant loop
contributions from the third--generation (s)fermions, charginos, $W^\pm$ and
charged Higgs bosons, are given by
\begin{eqnarray}
S^\gamma_i(\hat{s})&=&
2\sum_{f=\tau,b,t,\tilde{\chi}^\pm_1,\tilde{\chi}^\pm_2} N_C\, Q_f^2\,
g_fg^{S}_{H_i\bar{f}f}\,\frac{v}{m_f} F_{sf}(\tau_{f}) 
\nonumber \\ &&
- \sum_{\tilde{f}_j=\tilde{t}_1,\tilde{t}_2,\tilde{b}_1,\tilde{b}_2,
           \tilde{\tau}_1,\tilde{\tau}_2}
N_C\, Q_f^2g_{H_i\tilde{f}^*_j\tilde{f}_j}
\frac{v^2}{2m_{\tilde{f}_j}^2} F_0(\tau_{\tilde{f}_j})
\nonumber \\
&&- g_{_{H_iVV}}F_1(\tau_{W})-
g_{_{H_iH^+H^-}}\frac{v^2}{2 M_{H^\pm}^2} F_0(\tau_{H^\pm})
\,, \nonumber \\
P^\gamma_i(\hat{s})&=&2\sum_{f=\tau,b,t,\tilde{\chi}^\pm_1,\tilde{\chi}^\pm_2}
N_C\,Q_f^2\,g_fg^{P}_{H_i\bar{f}f}
\,\frac{v}{m_f} F_{pf}(\tau_{f})
 \,,
\end{eqnarray}
where $\tau_{x}=\hat{s}/4m_x^2$, $N_C=3$ for (s)quarks and $N_C=1$ for
(s)taus and charginos, respectively.
For the form factors $F_{sf}$, $F_{pf}$, $F_0$, and $F_1$ and the couplings
$g_f$, $g^{S,P}_{H_iff}$, $g_{H_i\tilde{f}^*_j\tilde{f}_j}$, $g_{_{H_iVV}}$,
and $g_{_{H_iH^+H^-}}$, we refer to Ref.~\refcite{CPsuperH}.

\smallskip

For numerical examples, we are going to consider two scenarios:
\begin{itemize}
\item \underline{CPX scenario}\cite{Carena:2000ks}
\begin{eqnarray}
&& \hspace{-3.0cm}
M_{\tilde{Q}_3} = M_{\tilde{U}_3} = M_{\tilde{D}_3} =
M_{\tilde{L}_3} = M_{\tilde{E}_3} = M_{\rm SUSY}\,,
\nonumber \\
&& \hspace{-3.0cm}
|\mu|=4\,M_{\rm SUSY}\,, \ \
|A_{t,b,\tau}|=2\,M_{\rm SUSY} \,, \ \
|M_3|=1 ~~{\rm TeV}\,.
\label{eq:CPX}
\end{eqnarray}
The parameter $\tan\beta$, the charged Higgs-boson pole mass
$M_{H^\pm}$, and the common SUSY scale $M_{\rm SUSY}$ can be varied.
For CP phases, taking $\Phi_\mu=0$ and a common
phase $\Phi_A=\Phi_{A_t}=\Phi_{A_b}=\Phi_{A_\tau}$ for $A$ terms, 
we have two
physical phases to vary: $\Phi_A$ and $\Phi_3={\rm Arg}(M_3)$.\\
 
\item \underline{Trimixing scenario}\cite{Ellis:2004fs}
\footnote{For an explicit example of 
the two nearly degenerate heavy Higgs bosons,
we refer to Ref.~\protect\refcite{Choi:2004kq}.}
\begin{eqnarray}
  \label{eq:Trimixing}
&&\tan\beta=50, \ \ M_{H^\pm}=155~~{\rm GeV},
\nonumber \\
&&M_{\tilde{Q}_3} = M_{\tilde{U}_3} = M_{\tilde{D}_3} =
M_{\tilde{L}_3} = M_{\tilde{E}_3} = M_{\rm SUSY} = 0.5 ~~{\rm TeV},
\nonumber \\
&& |\mu|=0.5 ~~{\rm TeV}, \ \
|A_{t,b,\tau}|=1 ~~{\rm TeV},   \ \
|M_2|=|M_1|=0.3~~{\rm TeV}, \ \ |M_3|=1 ~~{\rm TeV},
\nonumber \\
&&
\Phi_\mu = 0^\circ, \ \
\Phi_1 = \Phi_2 = 0^\circ\,,
\end{eqnarray}
and we have two varying phases: $\Phi_A$ and $\Phi_3$.
In this scenario, all the three-Higgs states mix significantly
with their widths larger than the mass differences.
For example, for $(\Phi_A,\Phi_3)=(90^\circ,-90^\circ)$, 
the code {\tt CPsuperH}\cite{CPsuperH} generates
\footnote{These values are slightly different from those given in 
Ref.~\protect\refcite{Ellis:2004fs} due to improvements
in the calculation of the neutral
Higgs-boson pole masses and the treatment of threshold corrections.}
\begin{eqnarray}
&&
M_{H_1}=118.3~~{\rm GeV}, \ \
M_{H_2}=119.9~~{\rm GeV}, \ \
M_{H_3}=123.6~~{\rm GeV}, \ \
\nonumber \\  &&
\Gamma_{H_1}=1.27~~{\rm GeV}, \ \ \ \
\Gamma_{H_2}=7.55~~{\rm GeV}, \ \ \ \ \ \
\Gamma_{H_3}=7.60~~{\rm GeV}.
\end{eqnarray}

\end{itemize}

\section{CP Violation in $s$-channel Production of Higgs Bosons}
\label{sec:schannel}

In the two-photon c.m. coordinate system with one photon momentum ${\bf k_1}$
along the positive $z$ direction and the other one ${\bf k_2}$ along the
negative $z$ direction, the wave vectors $\epsilon_{1,2}$ of two photons
are given by
\begin{eqnarray}
\epsilon_1(\lambda)=\epsilon^*_2(\lambda)
                   =\frac{1}{\sqrt{2}}\left(0,-\lambda,-i,0\right)\,,
\end{eqnarray}
where $\lambda=\pm 1$ denote the right and left photon helicities,
respectively.
Inserting the wave vectors into Eq.~(\ref{eq:hipp}) we obtain the production
helicity amplitude for the photon fusion process as follows:
\begin{eqnarray}
{\cal M}^{H_i}_{\lambda_1\lambda_2}=\frac{\alpha\,M_{H_i}^2}{4\pi\,v}
     \left\{ S^\gamma_i(M_{H_i}^2)+i \lambda_1 P^\gamma_i(M_{H_i}^2)\right\}
\delta_{\lambda_1\lambda_2}\,,
\label{hamp}
\end{eqnarray}
with $\lambda_{1,2}=\pm 1$, yielding the absolute polarized amplitude squared
\cite{Grzadkowski:1992sa,Choi:1999at}
\begin{eqnarray}
\overline{\left|{\cal M}^{H_i}\right|^2}=
           \overline{\left|{\cal M}^{H_i}\right|^2_0}\,\,
  \bigg\{             [1+\zeta_2\tilde{\zeta}_2]
      +{\cal A}_1\left[\zeta_2+\tilde{\zeta}_2\right]
      +{\cal A}_2\left[\zeta_1\tilde{\zeta}_3+\zeta_3\tilde{\zeta}_1\right]
      -{\cal A}_3\left[\zeta_1\tilde{\zeta}_1-\zeta_3\tilde{\zeta}_3\right]
                    \bigg\}\,,\nonumber \\
\label{mesq}
\end{eqnarray}
with the Stokes parameters $\{\zeta_i\}$ and $\{\tilde{\zeta}_i\}$ ($i=1,2,3$)
of two photon beams, respectively.
The first factor in Eq.~(\ref{mesq}) is the unpolarized amplitude squared;
\begin{eqnarray}
\overline{\left|{\cal M}^{H_i}\right|^2_0} =\frac{1}{4}
         \left\{\left|{\cal M}^{H_i}_{++}\right|^2
              +\left|{\cal M}^{H_i}_{--}\right|^2\right\}
=\frac{1}{2}\bigg\{\left|S^\gamma_i\right|^2+\left|P^\gamma_i\right|^2\bigg\}\,.
\end{eqnarray}
and three polarization asymmetries ${\cal A}^{H_i}_j$ ($j=1,2,3$) for each $H_i$
are defined in
terms of the helicity amplitudes and expressed in terms of the form factors
$S^\gamma_i$ and $P^\gamma_i$ as
\begin{eqnarray}
{\cal A}^{H_i}_1&=&
   \frac{\left|{\cal M}^{H_i}_{++}\right|^2-\left|{\cal M}^{H_i}_{--}\right|^2}
        {\left|{\cal M}^{H_i}_{++}\right|^2+\left|{\cal M}^{H_i}_{--}\right|^2}
    =\frac{2\,\imag(S^\gamma_iP^{\gamma_i\,*}_i)}
          {\left|S^\gamma_i\right|^2+\left|P^\gamma_i\right|^2}\,,
     \nonumber \\
{\cal A}^{H_i}_2&=&
   \frac{2\,\imag({\cal M}^{H_i\,*}_{--}{\cal M}^{H_i}_{++})}
        {\left|{\cal M}^{H_i}_{++}\right|^2+\left|{\cal M}^{H_i}_{--}\right|^2}
    =\frac{2\,\real(S^\gamma_iP^{\gamma_i\,*}_i)}
          {\left|S^\gamma_i\right|^2+\left|P^\gamma_i\right|^2}\,,
     \nonumber \\
{\cal A}^{H_i}_3&=&
   \frac{2\,\real({\cal M}^{H_i\,*}_{--}{\cal M}^{H_i}_{++})}
        {\left|{\cal M}^{H_i}_{++}\right|^2+\left|{\cal M}^{H_i}_{--}\right|^2}
    =\frac{\left|S^\gamma_i\right|^2-\left|P^\gamma_i\right|^2}
          {\left|S^\gamma_i\right|^2+\left|P^\gamma_i\right|^2}\,.
\end{eqnarray}
In the CP-invariant theories with real couplings, the form factors
$S^\gamma_i$ and $P^\gamma_i$ cannot exist simultaneously and
they should satisfy the relations;
${\cal A}^{H_i}_1={\cal A}^{H_i}_2=0$ and 
${\cal A}^{H_i}_3=+1(-1)$ depending on whether the
Higgs boson is a pure CP-even (CP-odd) state. In other words,
${\cal A}^{H_i}_1\neq 0$, ${\cal A}^{H_i}_2\neq 0$ and/or
$|{\cal A}^{H_i}_3|<1$ ensure that $H_i$ is 
a mixture of CP-even and CP-odd states, implying CP violation.
We note that both circularly and transversely polarized 
initial laser beams are needed to measure all three polarization asymmetries,
see Eqs.~(\ref{eq:zeta}) and (\ref{eq:Cs}).
Finally, the $s$-channel Higgs-boson production
$\gamma\gamma\rightarrow H_i$ is given by
\begin{equation}
\sigma(\gamma\gamma\rightarrow H_i)
=\hat{\sigma}_0(H_i)\,\delta(1-M_{H_i}^2/\hat{s})
 =\frac{\alpha^2}{32\pi v^2}
 \left(|S^\gamma_i|^2+|P^\gamma_i|^2\right)\delta(1-M_{H_i}^2/\hat{s})\,,
\label{eq:pph}
\end{equation}
where ${\alpha_{\rm em}^2}/{32\pi v^2}\sim 4$ fb.

\begin{figure}[htbh]
\vspace{-1.0cm}
\centerline{\epsfig{figure=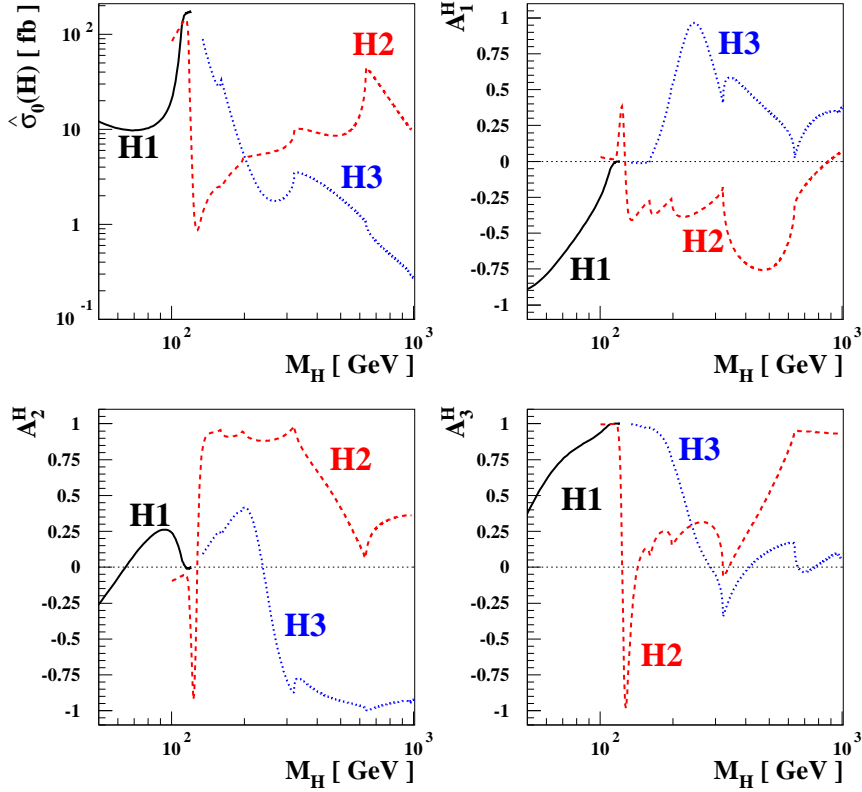,height=12cm,width=12cm}}
\vspace{-0.5cm}
\caption{{\it The parton cross-section $\hat{\sigma}_0(H_i)$ in units of fb and three
polarization asymmetries ${\cal A}_{1,2,3}^{H_i}$ as functions of 
each Higgs-boson mass for the CPX scenario taking
$\tan\beta=5$, $M_{\rm SUSY}=500$ GeV, and
$\Phi_A=\Phi_3=90^\circ$.
}}
\label{fig:xa_cpx}
\end{figure}

In Fig.~\ref{fig:xa_cpx}, we show the cross-section
$\hat{\sigma}_0(H_i)$ and three polarization asymmetries
${\cal A}_{1,2,3}^{H_i}$ as functions of
Higgs-boson masses taking the CPX scenario.
We observe sizeable effects of CP violation in all three
polarization asymmetries except ${\cal A}^{H_1}_{1,2,3}$ in
the decoupling limit $M_{H^\pm} \gsim ~200$ GeV.
Moreover, the polarization asymmetries ${\cal A}_1^{H_i}$ and
${\cal A}_2^{H_i}$ are complementary in the sense that 
one asymmetry is large where the other one is small.


\section{CP Violation in $\gamma\gamma\to \bar{f}f$}
\label{sec:pptoff}

In this section, we study the processes $\gamma\gamma\rightarrow f \bar{f}$
including the 
QED continuum background as well as the Higgs contribution.
\cite{Asakawa:1999gz,Asakawa:2000jy,Godbole:2002qu,Ellis:2004hw,Godbole:2006eb}

\subsection{Helicity Amplitudes}
%
\begin{figure}[tbh]
\begin{center}
\begin{picture}(300,100)(0,0)

\Photon(0,90)(60,90){3}{5}
\Photon(0,10)(60,10){3}{5}
\ArrowLine(60,90)(120,90)
\ArrowLine(60,10)(60,90)
\ArrowLine(120,10)(60,10)

\Text(-10,90)[]{$\gamma$}
\Text(60,100)[]{$\mu$}
\ArrowLine(20,80)(40,80)
\Text(30,70)[]{$k_1,\epsilon_1(\lambda_1)$}

\Text(-10,10)[]{$\gamma$}
\Text(60,0)[]{$\nu$}
\ArrowLine(20,20)(40,20)
\Text(30,30)[]{$k_2,\epsilon_2(\lambda_2)$}

\Text(130,93)[]{$f$}
\ArrowLine(80,95)(100,95)
\Text(90,105)[]{$p_1,\sigma$}

\Text(130,13)[]{$\bar{f}$}
\ArrowLine(80,15)(100,15)
\Text(90,25)[]{$p_2,\bar{\sigma}$}

\Photon(180,10)(240,90){3}{5}
\Photon(180,90)(240,10){3}{5}
\ArrowLine(240,90)(300,90)
\ArrowLine(240,10)(240,90)
\ArrowLine(300,10)(240,10)

\Text(240,100)[]{$\nu$}
\Text(240,0)[]{$\mu$}
\end{picture}\\
\end{center}
\smallskip
\noindent
\caption{\it Feynman diagrams contributing to the tree-level QED
background, introducing our definitions of the initial-state photon and
final-state fermion momenta and helicities.}
\label{fig:QED}
\end{figure}
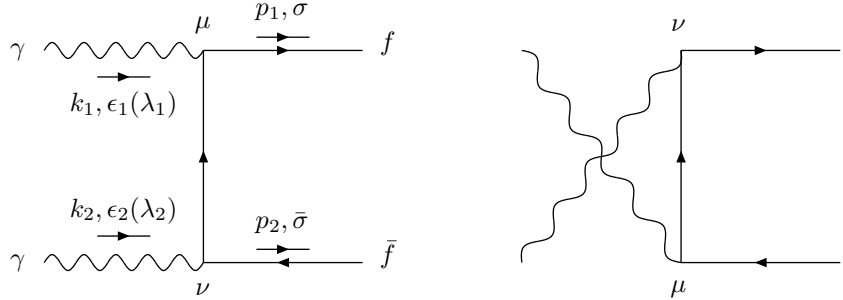

The tree-level Feynman diagrams for the QED process $\gamma \gamma
\rightarrow {\bar f} f$ are shown in Fig.~\ref{fig:QED}.
In the two-photon c.m. system, the helicity amplitudes
for the QED production of a fermion-antifermion pair take the forms:
\begin{eqnarray}
{\cal M}_C &=& 4\pi\alpha Q_f^2  \
\langle \sigma\,\bar{\sigma};\lambda_1\,\lambda_2 \rangle_C ,
\label{eq:MC}
\end{eqnarray}
where
\begin{eqnarray}
\langle \sigma\,\sigma;\lambda\,\lambda \rangle_C &=&
\frac{4 m_f}{\sqrt{\hat{s}}}\
\frac{1}{1-\beta_f^2 c_\theta^2} \ (\lambda+\sigma \beta_f)\,; \ \ \
\langle \sigma\,\sigma;\lambda\,-\lambda \rangle_C =
-\frac{4 m_f}{\sqrt{\hat{s}}}\
\frac{s_\theta^2}{1-\beta_f^2 c_\theta^2} \ \sigma \beta_f\,;
\nonumber \\
\langle \sigma\,-\sigma;\lambda\,\lambda \rangle_C &=& 0\,; \ \ \
\langle \sigma\,-\sigma;\lambda\,-\lambda \rangle_C =
-2 \beta_f \
\frac{s_\theta}{1-\beta_f^2 c_\theta^2} \ (\sigma\lambda+c_\theta) \,.
\label{eq:QEDhel}
\end{eqnarray}
We use the abbreviations
$s_\theta\equiv \sin\theta$ and $c_\theta\equiv \cos\theta$
with $\theta$ the angle between ${\bf p_1}$ and ${\bf k_1}$,
and $\beta_f \equiv \sqrt{1-4m_f^2/\hat{s}}$ with
$\hat{s} = (k_1+k_2)^2 = (p_1+p_2)^2$.
We allow for independent and measurable polarizations $\lambda_{1,2}$ of the
initial-state photons and ${\bar \sigma}, \sigma$ of
the final-state fermion-antifermion pair. 
We note that the last amplitude
in (\ref{eq:QEDhel}) with completely different
helicity states is the least important, since the Higgs-mediated diagram
is nonvanishing only when the helicities of photons and/or those of
final fermions are equal.

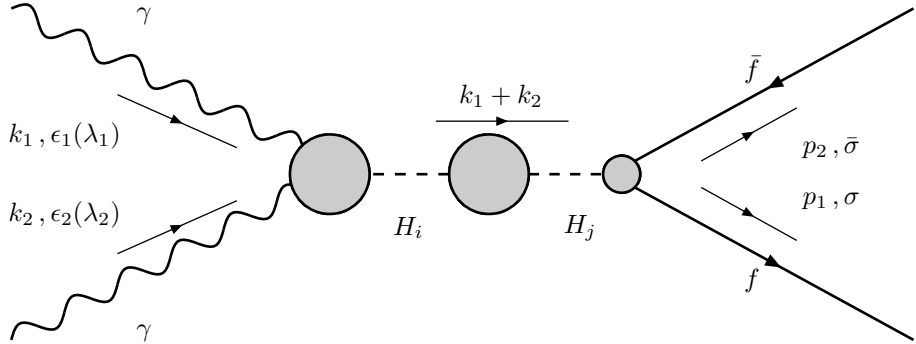
\begin{figure}[tbh]
\vspace*{1cm}
\begin{center}

\SetWidth{1.0}

\begin{picture}(250,100)(0,0)
\Photon(-50,112.5)(70,55){4}{6}
\Photon(-50,-12.5)(70,50){4}{6}
\DashCArc(130,50)(15,0,360){4}
\GOval(130,50)(15,15)(0){0.8}
\DashLine(70,50)(115,50){4}
\DashCArc(70,50)(15,0,360){4}
\GOval(70,50)(15,15)(0){0.8}
\DashLine(145,50)(180,50){4}
\DashCArc(180,50)(7,0,360){4}
\GOval(180,50)(7,7)(0){0.8}
\ArrowLine(290,112.5)(185,55)
\ArrowLine(185,45)(290,-12.5)
\Text(0,110)[]{$\gamma$}
\Text(0,-10)[]{$\gamma$}
\Text(100,30)[]{$H_i$}
\Text(165,30)[]{$H_j$}
\Text(230,90)[]{$\bar{f}$}
\Text(230,10)[]{$f$}
\SetWidth{0.5}
\ArrowLine(210,55)(246,75)
\ArrowLine(210,45)(246,25)
\Text(260,60)[]{$p_2\,,\bar{\sigma}$}
\Text(260,40)[]{$p_1\,,\sigma$}
\ArrowLine(-10,20)(35,40)
\ArrowLine(-10,80)(35,60)
\Text(-30,35)[]{$k_2\,,\epsilon_2 (\lambda_2)$}
\Text(-30,65)[]{$k_1\,,\epsilon_1 (\lambda_1)$}
\ArrowLine(110,70)(160,70)
\Text(135,80)[]{$k_1+k_2$}
\end{picture}
\end{center}
\smallskip
\noindent
\vspace{0.2cm}
\caption{\it Mechanisms contributing to the process $\gamma \gamma
\rightarrow H_{1,2,3} \to {\bar f} f$, including off-diagonal
absorptive parts in the Higgs-boson propagator matrix.}
\label{fig:Higgs}
\end{figure}
 
The helicity amplitudes of the process
$\gamma \gamma \rightarrow H \rightarrow {\bar f} f$, see
Fig.~\ref{fig:Higgs}, are given by
\begin{equation}
{\cal M}_H \; = \; \frac{\alpha \, m_f \, \sqrt{\hat{s}}}{4\pi v^2}
\langle \sigma ;\lambda_1 \rangle_H
\delta_{\sigma\bar{\sigma}}\delta_{\lambda_1\lambda_2},
\label{eq:MH}
\end{equation}
where the reduced amplitude
\footnote{Here we have corrected an error in the overall sign of the amplitude.
\cite{Ellis:2004hw}}
\begin{equation}
\langle \sigma ;\lambda \rangle_H \; = \; -\sum_{i,j=1}^3
[S_i^\gamma(\hat{s})+i\lambda P_i^\gamma(\hat{s})]
\ D_{ij}(\hat{s}) \
(\sigma\beta_f g^S_{H_j\bar{f}f}-ig^P_{H_j\bar{f}f}),
\label{eq:redHamp}
\end{equation}
is a quantity given by the Higgs-boson propagator matrix $D_{ij}(\hat{s})$
combined with the production and decay vertices. 
In the calculation of the propagator matrix $D_{ij}(\hat{s})$, 
we have fully included the off-diagonal absorptive parts which cannot be 
neglected when two or more MSSM Higgs bosons contribute
simultaneously to the process under consideration.
\cite{Ellis:2004fs,Lee:2004wx}
%

We note the following properties of the $\gamma
\gamma
\rightarrow
{\bar f} f$ helicity amplitudes under the CP transformation:
\begin{equation}
\langle \sigma\,\bar{\sigma};\lambda_1\,\lambda_2 \rangle  \ \
\stackrel{\rm CP}{\leftrightarrow} \ \
(-1)(-1)^{(\sigma-\bar{\sigma})/2}
\langle -\bar{\sigma}\,-\sigma;-\lambda_2\,-\lambda_1 \rangle  .
\label{eq:cp}
\end{equation}
Also interesting are the properties under the CP$\widetilde{\rm T}$
transformation, where $\widetilde{\rm T}$ reverses the signs of the spins
and the three-momenta of the
asymptotic states, without interchanging initial and final
states, and the matrix element gets complex conjugated:
\begin{equation}
\langle \sigma\,\bar{\sigma};\lambda_1\,\lambda_2 \rangle  \ \
\stackrel{\rm CP\widetilde{\rm T}}{\leftrightarrow} \ \
(-1)(-1)^{(\sigma-\bar{\sigma})/2}
\langle -\bar{\sigma}\,-\sigma;-\lambda_2\,-\lambda_1 \rangle^*  .
\label{eq:cpt}
\end{equation}
Evidently, the QED helicity amplitudes (\ref{eq:QEDhel}) are even under
both the CP and CP$\widetilde{\rm T}$ transformations.
On the other hand, the simultaneous presence of
$\{S_i^\gamma, P_i^\gamma\}$ and/or $\{g^S_{H_j\bar{f}f}, g^P_{H_j\bar{f}f}\}$
would signal CP violation in the Higgs-boson-exchange amplitude
(\ref{eq:redHamp}), and nonvanishing absorptive parts from the vertices
and the propagators could also lead to
CP$\widetilde{\rm T}$ violation in the Higgs-exchange diagram.


\subsection{Identical Photon and Fermion Helicities}

The most appealing case may be that of identical photon and fermion helicities due to
the $m_f$ suppression of the large QED background (\ref{eq:QEDhel})
and the relative easiness in 
controlling the mean helicities of the colliding photon beams and 
measuring the longitudinal polarizations of the final fermions.
In this case the amplitude may be written as
\begin{eqnarray}
{\cal M}^{\rm I}_{\sigma\lambda}=
\left. {\cal M}_C \right|_{\bar{\sigma}=\sigma, \lambda_1=\lambda_2=\lambda}
+{\cal M}_H
&=& \frac{\alpha \, m_f \, \sqrt{\hat{s}}}{4\pi v^2}
\langle \sigma ;\lambda \rangle
\end{eqnarray}
where ${\cal M}_C$ and ${\cal M}_H$ are given in
Eqs.~(\ref{eq:MC}) and (\ref{eq:MH}), respectively.
The amplitude $\langle \sigma ;\lambda \rangle$ consists of two terms as
\begin{equation}
\langle \sigma ;\lambda \rangle \; \equiv \;
 \langle \sigma ;\lambda \rangle_H + R(\hat{s}) f(\theta)
\langle \sigma ;\lambda \rangle_C \,,
\label{eq:helamp}
\end{equation}
where $R(\hat{s}) = 64\pi^2 Q_f^2\, v^2/\hat{s}$,
$f(\theta) = 1/(1-\beta_f^2 c_\theta^2)$, and
$\langle \sigma ;\lambda \rangle_C  = \lambda+\sigma\beta_f$, see Eq.~(\ref{eq:QEDhel}).
We note that $\langle \sigma=\mp ;\lambda=\pm \rangle_C  = \pm(1-\beta_f)$, which
vanishes in the limit $m_f \rightarrow 0$.

Depending on the different combinations of helicities of the
initial-state photons and final-state fermions, we have four differential
cross-sections:
\begin{equation}
\frac{{\rm d}\hat{\sigma}_{\sigma\lambda}}{{\rm d}c_\theta}
=\frac{\beta_f N_C}{32 \pi \hat{s}}
\left|{\cal M}^{\rm I}_{\sigma\lambda}\right|^2
\end{equation}
where $\sigma\,,\lambda=\pm\,.$ After integrating them over $c_\theta$, we have
the cross-sections
\begin{equation}
\hat{\sigma}_{\sigma\lambda}
=\frac{\beta_f N_C}{32 \pi}
\left(\frac{\alpha \, m_f }{4\pi v^2} \right)^2
{\cal Y}_{\sigma\lambda}
\label{eq:cx1}
\end{equation}
with the $\hat{s}$-dependent functions
\begin{eqnarray}
{\cal Y}_{\sigma\lambda}&\equiv &
\int_{-z_f}^{z_f} {\rm d}c_\theta \left|\langle\sigma;\lambda\rangle\right|^2
\\
&=&2\bigg\{z_f\left|\langle \sigma;\lambda \rangle_H \right|^2
+R(\hat{s})^2\, F_1^{z_f} \left|\langle \sigma;\lambda \rangle_C \right|^2
+R(\hat{s})\, F_2^{z_f} \ 
\real\left[\langle \sigma;\lambda \rangle_H
\langle \sigma;\lambda \rangle_C^*\right] \bigg\}\,. \nonumber
\label{eq:y}
\end{eqnarray}
The functions $F_1^{z_f}$ and $F_2^{z_f}$ are given by
\begin{eqnarray}
F_1^{z_f} &=&\frac{1}{2}\int_{-z_f}^{z_f}{\rm d}c_\theta f^2(\theta)=
\frac{z_f}{2(1-z_f^2\beta_f^2)}+
\frac{\ln{\frac{1+z_f\beta_f}{1-z_f\beta_f}}}{4\beta_f}\,,
\nonumber \\
F_2^{z_f} &=&\int_{-z_f}^{z_f}{\rm d}c_\theta f(\theta)
=\frac{\ln{\frac{1+z_f\beta_f}{1-z_f\beta_f}}}{\beta_f} \,.
\label{eq:f12}
\end{eqnarray}
Note that we have introduced an experimental cut on the fermion polar angle
$\theta$ : $|\cos\theta| \leq z_f$ and $\cos\theta^f_{\rm cut}=z_f$.
Experimentally, we cannot measure the final state
fermion if it has too small angle $\theta$ outside the coverage of detectors.
This angular cut has significant effects in the cases of light fermions,
$f=\mu$, $\tau$, and $b$, since the QED continuum differential cross-section
${\rm d}\hat{\sigma}_C/{\rm d}c_\theta$ is peaked 
in the forward and backward directions. 
This makes the
cross-section $\hat{\sigma}_C$, or $F_1^{z_f}$, strongly depend on
$z_f$ when the final fermions are light:
\begin{equation}
\hat{\sigma}_C\equiv\frac{\beta_f N_C}{16 \pi}
\left(\frac{\alpha \, m_f }{4\pi v^2} \right)^2
\,R(\hat{s})^2\,F^{z_f}_1\,
\frac{\sum_{\sigma,\lambda=\pm}
\left|\langle\sigma;\lambda\rangle_C\right|^2}{4}\,.
\end{equation}
Actually the QED cross-sections are suppressed by
factors of about 5000 and 20 for $f=\mu$ and $f=\tau$ cases, respectively,
by imposing $\theta^{\mu,\tau}_{\rm cut}=130$ mrad angle cut
($z_{\mu,\tau}\simeq 0.99$) when $\sqrt{\hat{s}}=120$ GeV.
For $b$-quark case, the suppression factor is about 30 imposing
$\theta^b_{\rm cut}=280$ mrad ($z_b\simeq 0.96$).
On the other hand, the Higgs-mediated cross-section and the QED continuum
cross-section for top quarks are hardly affected by the polar angle cut.
The introduction of the polar angle cut, therefore, greatly enhance the
significance of the Higgs-mediated process with respect to the QED continuum
one for $f=\mu\,,\tau\,,$ and $b$.

The CP and CP$\widetilde{\rm T}$ parities of the polarization-dependent cross
sections $\hat{\sigma}_{\sigma\lambda}$ (\ref{eq:cx1})
can easily be obtained by observing that
\begin{equation}
{\cal Y}_{\sigma\lambda} \ \stackrel{\rm CP}{\leftrightarrow}
\ {\cal Y}_{-\sigma-\lambda}\,, \ \
{\cal Y}_{\sigma\lambda} \stackrel{\rm CP\widetilde{\rm T}}{\leftrightarrow}
{\cal Y}_{-\sigma-\lambda}\,, \ \
\end{equation}
which are derived from (\ref{eq:cp}) and (\ref{eq:cpt}). 
Based on this observation,
we can construct two CP-violating cross-sections in
terms of $\hat{\sigma}_{\sigma\lambda}$:
\begin{equation}
\hat\Delta_1 \equiv \hat{\sigma}_{++}-\hat{\sigma}_{--}\,, \ \
\hat\Delta_2 \equiv \hat{\sigma}_{+-}-\hat{\sigma}_{-+}\,,
\end{equation}
or, equivalently, the two linear combinations
\begin{equation}
(\hat\Delta_1  + \hat\Delta_2) = \sum_{\lambda=\pm}
(\hat{\sigma}_{+\lambda}-\hat{\sigma}_{-\lambda})\,, \ \
(\hat\Delta_1  - \hat\Delta_2) = \sum_{\sigma=\pm}
(\hat{\sigma}_{\sigma +}-\hat{\sigma}_{\sigma -})\,.
\end{equation}
Note that the CP-violating cross-section ($\hat\Delta_1-\hat\Delta_2$) can be
determined without measuring the helicities of the final-state fermions.
Finally, the unpolarized total cross-section is given by
\begin{equation}
\hat{\sigma}_{\rm tot}=\frac{1}{4}\left(
\hat{\sigma}_{++}+\hat{\sigma}_{--}+
\hat{\sigma}_{+-}+\hat{\sigma}_{-+} \right) \,.
\end{equation}

\begin{figure}[htbh]
\vspace{-1.0cm}
\centerline{\epsfig{figure=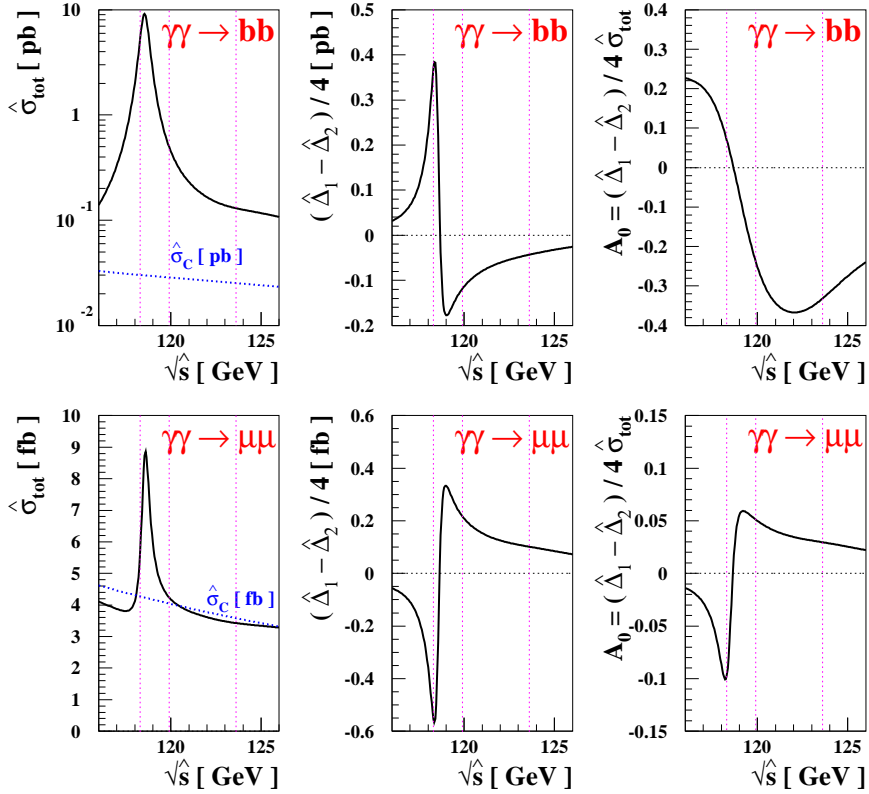,height=12cm,width=12cm}}
\vspace{-0.5cm}
\caption{{\it The cross-sections $\hat{\sigma}_{\rm tot}$ and
$(\hat{\Delta}_1-\hat{\Delta}_2)/4$ and the CP asymmetry ${\cal A}_0$ 
as functions of $\sqrt{\hat{s}}$
for the processes $\gamma\gamma\rightarrow \bar{b}b$ (upper frames) and
$\gamma\gamma\rightarrow \mu^+\mu^-$ (lower frames) in the trimixing scenario with
$(\Phi_A,\Phi_3)=(90^\circ,-90^\circ)$. The continuum cross-sections $\hat{\sigma}_C$ are
also shown in the left frames as dotted lines. The three Higgs masses are indicated by
vertical lines.
}}
\label{fig:bmuon}
\end{figure}

First we consider the processes $\gamma\gamma\rightarrow \bar{b}b$ and
$\gamma\gamma\rightarrow \mu^+\mu^-$ in which the 
helicities of the final-state fermions cannot be measured.
In Fig.~\ref{fig:bmuon}, we show the 
 cross-sections $\hat{\sigma}_{\rm tot}$ and
$(\hat{\Delta}_1-\hat{\Delta}_2)/4$ and the  CP asymmetry ${\cal A}_0$ defined by
\begin{equation}
{\cal A}_0\equiv \frac{\hat{\Delta}_1-\hat{\Delta}_2}{4\,\hat{\sigma}_{\rm tot}}\,.
\end{equation}
Assuming an integrated $\gamma\gamma$ luminosity of
100 fb$^{-1}$, and high efficiency for $b$-quark reconstruction, we may
expect more than ten thousand events with the total cross-section 
larger than $\sim$ 0.1 pb in the process
$\gamma\gamma \rightarrow \bar{b}b$.
This would enable
one to probe CP asymmetry at the 1 \% level or less.
For the process $\gamma\gamma\rightarrow \mu^+\mu^-$
we expect to have a few hundred events 
with a 100 fb$^{-1}$ integrated $\gamma\gamma$ luminosity.
The CP asymmetry ${\cal A}_0$
can be probed only when it is larger than a few \%.
But, differently from the $\gamma\gamma \rightarrow \bar{b}b$ process,
the good resolution in the muon invariant mass
enables us to examine
the $\sqrt{\hat{s}}$ dependence of the cross-sections and the CP asymmetry.

\begin{figure}[htbh]
\vspace{-1.0cm}
\centerline{\epsfig{figure=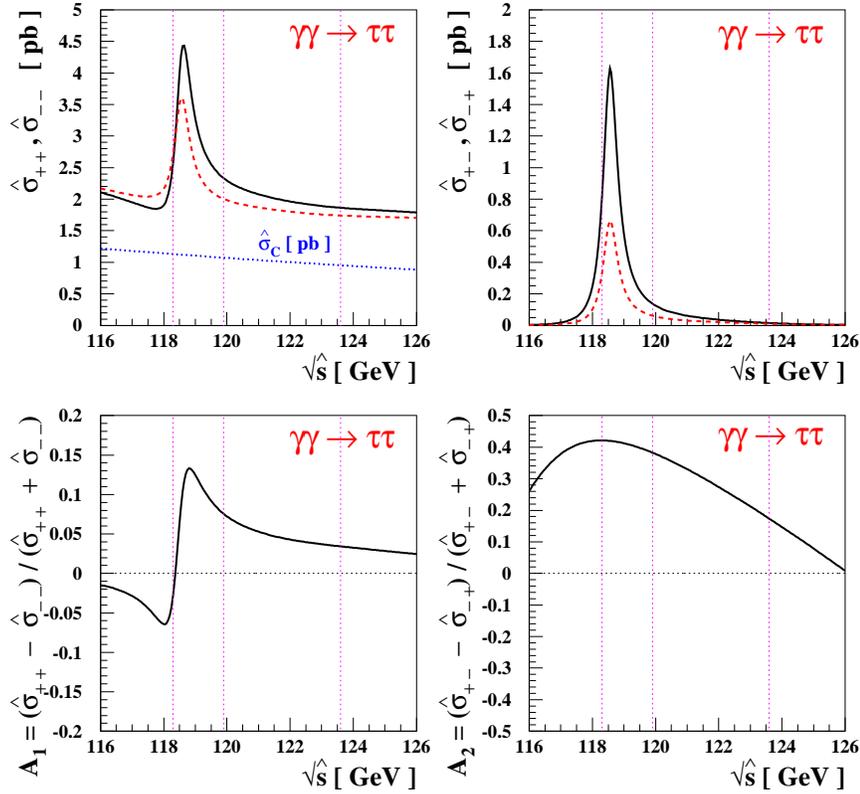,height=12cm,width=12cm}}
\vspace{-0.5cm}
\caption{{\it The cross-sections $\hat{\sigma}_{\sigma\lambda}$ (upper frames) and
the CP asymmetries ${\cal A}_{1,2}$ (lower frames)
as functions of $\sqrt{\hat{s}}$
for the processes $\gamma\gamma\rightarrow \tau^+\tau^-$ 
in the trimixing scenario with
$(\Phi_A,\Phi_3)=(90^\circ,-90^\circ)$. 
The solid lines are for $\hat{\sigma}_{++}$ and $\hat{\sigma}_{+-}$ and
the dashed lines for $\hat{\sigma}_{--}$ and $\hat{\sigma}_{-+}$ in the upper two frames.
The continuum cross-section $\hat{\sigma}_C$ is
also shown in the upper-left frame as a dotted line. The three Higgs masses are indicated by
vertical lines.
}}
\label{fig:tau}
\end{figure}

In the process $\gamma\gamma\rightarrow \tau^+\tau^-$, the polarizations of final tau
leptons can be measured through its decays. 
In Fig.~\ref{fig:tau}, we show the cross-sections
$\hat{\sigma}_{\sigma\lambda}$ in (\ref{eq:cx1}) in the upper frames. 
When $\sigma=\lambda$ (upper-left frame), the cross-sections are around 2-4 pb
and we observe the sizable difference
between $\hat{\sigma}_{++}$ (solid) and $\hat{\sigma}_{--}$ (dashed), which is just the
$\hat\Delta_1$.
When $\sigma\neq\lambda$ (upper-right frame), the cross-sections are
peaked between $M_{H_1}$ and $M_{H_2}$
with their sizes of about 1.6 pb ($\hat{\sigma}_{+-}$) and
0.6 pb ($\hat{\sigma}_{-+}$) as shown by solid and dashed lines,
respectively.  Again we see a sizable difference
between the two cross-sections, which is the CP-violating cross-section
$\hat\Delta_2$.
One may define two CP-odd asymmetries as:
\begin{equation}
{\cal A}_1\equiv
\frac{\hat\Delta_1}{\hat{\sigma}_{++}+\hat{\sigma}_{--}}\,, \ \ \ \
{\cal A}_2\equiv
\frac{\hat\Delta_2}{\hat{\sigma}_{+-}+\hat{\sigma}_{-+}}\,.
\label{eq:cpasym}
\end{equation}
Assuming 10,000 $\gamma\gamma\to\tau^+\tau^-$ events after applying
some experimental cuts to reconstruct the tau leptons and their polarizations, CP-odd
asymmetries larger than 1 \% could be measured.

\subsection{Initial and Final Spin-spin Correlations}
The polarizations of the final fermions can be measured
in the processes such as 
$\gamma\gamma\rightarrow \tau^+\tau^-$, $\bar{t}t$, $\chi^+_i\chi^-_j$, 
$\chi^0_i\chi^0_j$, etc. 
In this case, one can construct a lot of observables
through initial and final spin-spin correlations.
It is convenient to consider two cases separately:
$(i)$ identical fermion helicities, $\sigma=\bar{\sigma}$ and
$(ii)$ identical photon helicities, $\lambda=\lambda_1=\lambda_2$.

\subsubsection{Identical fermion helicities: initial photon spin-spin correlations}

In this case, for each $\sigma$, the amplitude can be cast into a $2\times 2$
matrix:
\begin{equation}
\left({\cal M}^{\rm II}_\sigma\right)_{\lambda_1\lambda_2}=
\frac{\alpha \, m_f \, \sqrt{\hat{s}}}{4\pi v^2} \left(
\begin{array}{cc}
\langle \sigma;+ \rangle & \langle\sigma\rangle_C  \\
\langle\sigma\rangle_C & \langle \sigma;- \rangle
\end{array} \right)
\ \ {\rm with} \ \
\langle \sigma \rangle_C \equiv -R(\hat{s})\,s_\theta^2\,f(\theta)\,
\sigma\beta_f \,,
\end{equation}
and the
$\langle \sigma;\lambda \rangle $ defined in (\ref{eq:helamp}).
With the polarization density matrices for the two photons:
\begin{eqnarray}
\tilde{\rho}=\frac{1}{2}\left(\begin{array}{cc}
                   1+\tilde{\zeta}_2 & -\tilde{\zeta}_3+i\tilde{\zeta}_1 \\
                   -\tilde{\zeta}_3-i\tilde{\zeta}_1 & 1-\tilde{\zeta}_2\\
                   \end{array}\right) \,, \ \
\rho=\frac{1}{2}\left(\begin{array}{cc}
                   1+\zeta_2 & -\zeta_3+i\zeta_1 \\
                   -\zeta_3-i\zeta_1 & 1-\zeta_2\\
                   \end{array}\right) \,, \ \
\end{eqnarray}
the polarization-weighted squared matrix elements can be obtained by
\cite{Hagiwara:1985yu}
\begin{equation}
\left|{\cal M}^{\rm II}_\sigma\right|^2
={\rm Tr} \left[ {\cal M}^{\rm II}_\sigma \, \tilde{\rho} \,
 {\cal M}^{\rm II\,\dagger}_\sigma \, \rho^T \right] \,.
\end{equation}
The amplitude squared for each  $\sigma$ can be expanded as
\begin{eqnarray}
\left|{\cal M}^{\rm II}_\sigma\right|^2 &=&
\left(\frac{\alpha m_f \sqrt{\hat{s}}}{4\pi v^2}\right)^2  
\Bigg\{
A^\sigma_1(1+\zeta_2\tilde{\zeta}_2)+
A^\sigma_2[(\zeta_1\tilde{\zeta}_1-\zeta_2\tilde{\zeta}_2)+
(\zeta_3\tilde{\zeta}_3-\zeta_2\tilde{\zeta}_2)]
\nonumber \\ &&
+B^\sigma_1(\zeta_2+\tilde{\zeta}_2)
+B^\sigma_2(\zeta_1\tilde{\zeta}_3+\zeta_3\tilde{\zeta}_1)
+B^\sigma_3(\zeta_3\tilde{\zeta}_3-\zeta_1\tilde{\zeta}_1)
\nonumber \\ &&
+C^\sigma_1(\zeta_1+\tilde{\zeta}_1)
+C^\sigma_2(\zeta_3+\tilde{\zeta}_3)
+C^\sigma_3(\zeta_1\tilde{\zeta}_2+\zeta_2\tilde{\zeta}_1)
+C^\sigma_4(\zeta_2\tilde{\zeta}_3+\zeta_3\tilde{\zeta}_2)
\Bigg\} \,, \nonumber
\end{eqnarray}
where the polarization coefficients are given by
\begin{eqnarray}
A^\sigma_1 &=& \frac{1}{4}\left(|\langle \sigma;+ \rangle|^2 +|\langle
\sigma;- \rangle|^2 +2\,|\langle \sigma \rangle_C|^2\right) \,; \ 
A^\sigma_2 = \frac{1}{2}\,|\langle \sigma \rangle_C|^2\,, \nonumber \\
B^\sigma_1 &=& \frac{1}{4}\left(|\langle \sigma;+ \rangle|^2 -|\langle
\sigma;- \rangle|^2\right) \,; \ 
B^\sigma_2 = 
\frac{1}{2}\,\imag[\langle \sigma;+ \rangle\,\langle \sigma;- \rangle^*] \,;
\nonumber \\
B^\sigma_3 &=& \frac{1}{2}\,\real[\langle \sigma;+ \rangle\,\langle \sigma;- \rangle^*]
\,, \nonumber \\
C^\sigma_1 &=& -\frac{1}{2}\,\imag[(\langle \sigma;+ \rangle -
\langle \sigma;- \rangle)\,
\langle \sigma \rangle_C^*] \,;  \
C^\sigma_2 = -\frac{1}{2}\,\real[(\langle \sigma;+ \rangle +
\langle \sigma;- \rangle)\,
\langle \sigma \rangle_C^*] \,; \nonumber \\
C^\sigma_3 &=& -\frac{1}{2}\,\imag[(\langle \sigma;+ \rangle +
\langle \sigma;- \rangle)\,
\langle \sigma \rangle_C^*] \,;  \
C^\sigma_4 = -\frac{1}{2}\,\real[(\langle \sigma;+ \rangle -
\langle \sigma;- \rangle)\,
\langle \sigma \rangle_C^*] \,.
\end{eqnarray}
We note that $B^\sigma_{2,3}$ are related to the
observables in the
interference between the amplitudes with different photon helicities requiring
linear polarizations of the colliding photon beams, and that
the observables $C^\sigma_i$ are due to interference with the QED
continuum.

\subsubsection{Identical photon helicities: final fermion spin-spin correlations}

In this case, for each $\lambda$, the amplitude can be cast into
a diagonal $2\times 2$ matrix:
\begin{equation}
\left({\cal M}^{\rm III}_\lambda\right)_{\sigma\bar{\sigma}}=
\frac{\alpha \, m_f \, \sqrt{\hat{s}}}{4\pi v^2} \left(
\begin{array}{cc}
\langle +;\lambda \rangle & 0  \\
0 & \langle -;\lambda \rangle
\end{array} \right) \,,
\end{equation}
where $\langle \pm;\lambda \rangle $ are defined in (\ref{eq:helamp}).
The polarization density matrices for the two final-state fermions are
\begin{eqnarray}
\bar{\rho}=\frac{1}{2}\left(\begin{array}{cc}
                   1+\bar{P}_L & -\bar{P}_T e^{i\bar{\alpha}} \\
                   -\bar{P}_T e^{-i\bar{\alpha}} & 1-\bar{P}_L\\
                   \end{array}\right) \,, \ \
\rho=\frac{1}{2}\left(\begin{array}{cc}
                   1+P_L & P_T e^{-i\alpha} \\
                   P_T e^{i\alpha} & 1-P_L\\
                   \end{array}\right) \,. \ \
\end{eqnarray}
Here, $P_L$ and $\bar{P}_L$ are the longitudinal polarizations of the
fermion $f$ and antifermion $\bar{f}$, respectively,
while $P_T$ and $\bar{P}_T$ are the degrees
of transverse polarization with $\alpha$ and $\bar{\alpha}$ being
the azimuthal angles with respect to the production plane.
For each $\lambda$, the polarization-weighted squared matrix elements are
\cite{Hagiwara:1985yu}
\begin{eqnarray}
\left|{\cal M}^{\rm III}_\lambda\right|^2
={\rm Tr} \left[ {\cal M}^{\rm III}_\lambda \, \bar{\rho}^T \,
 {\cal M}^{\rm III\,\dagger}_\lambda \, \rho \right] 
&=&
\left(\frac{\alpha m_f \sqrt{\hat{s}}}{4\pi v^2}\right)^2
\Bigg\{
D^\lambda_1 (1+P_L\bar{P}_L)
+D^\lambda_2 (P_L+\bar{P}_L)
\nonumber \\
&& 
+P_T\bar{P}_T[D^\lambda_3\cos(\alpha-\bar{\alpha})
+D^\lambda_4\sin(\alpha-\bar{\alpha})]
\Bigg\} \,,
\end{eqnarray}
where
\begin{eqnarray}
D_1^\lambda &=& \frac{1}{4}
\left(|\langle +;\lambda \rangle|^2 +|\langle -;\lambda \rangle|^2\right)
\,; \
D_2^\lambda = \frac{1}{4}
\left(|\langle +;\lambda \rangle|^2 -|\langle -;\lambda \rangle|^2\right)
\,; \nonumber \\
D_3^\lambda &=& -\frac{1}{2}
\real\left(\langle +;\lambda \rangle \langle -;\lambda \rangle^*\right)
\,; \
D_4^\lambda = \frac{1}{2}
\imag\left(\langle +;\lambda \rangle \langle -;\lambda \rangle^*\right)
\,.
\end{eqnarray}
The quantities $D^\lambda_{3,4}$ are related to the observables coming
from the interference between the amplitudes with different fermion
helicities, for which we need to determine the transverse polarizations of
the final fermions.

\subsubsection{Observables}
We consider two categories of cross-sections, according to the
final-state fermion helicity $\sigma$ or 
the initial-state photon helicity $\lambda$:
\begin{eqnarray}
\frac{d\hat\Sigma^X}{dc_\theta} &\equiv &
\frac{\beta_f N_C}{32 \pi }
\left(\frac{\alpha m_f }{4\pi v^2}\right)^2 (X^++X^-)\,,
\nonumber \\
\frac{d\hat\Delta^X}{dc_\theta} &\equiv &
\frac{\beta_f N_C}{32 \pi}
\left(\frac{\alpha m_f}{4\pi v^2}\right)^2 (X^+-X^-)\,,
\end{eqnarray}
where $X^\pm=A^\sigma_i,B^\sigma_j,C^\sigma_k,D^\lambda_l$.
In this way, we can construct more than 20 independent observables. It can be shown
that half of the observables
are CP-odd using Eqs.~(\ref{eq:cp}) and  (\ref{eq:cpt}), 
see Table~\ref{tab:tab}.
We refer to Ref.~\refcite{Ellis:2004hw} for some numerical examples
of the observables $\hat\Sigma^X$ and $\hat\Delta^X$ in the processes
$\gamma\gamma\rightarrow \tau^+\tau^-$ and 
$\gamma\gamma\rightarrow \bar{t}t$.

\begin{table}[h]
\tbl{\it The {\rm CP} and {\rm CP}$\widetilde{\rm T}$ parities of the cross-sections.}
{\begin{tabular}{@{}cccc@{}} \toprule
$A$ type $[{\rm CP},{\rm CP\tilde{T}}]$ & $B$ type $[{\rm CP},{\rm CP\tilde{T}}]$ & 
$C$ type $[{\rm CP},{\rm CP\tilde{T}}]$ & $D$ type $[{\rm CP},{\rm CP\tilde{T}}]$ \\
\colrule
$\hat{\Sigma}^{A_1}[+,+]$ & $\hat{\Sigma}^{B_1}[-,-]$ &
$\hat{\Sigma}^{C_1}[-,+]$ & $\hat{\Sigma}^{D_1}[+,+]$ \\
$\hat{\Sigma}^{A_2}[+,+]$ & $\hat{\Sigma}^{B_2}[-,+]$ &
$\hat{\Sigma}^{C_2}[+,+]$ & $\hat{\Sigma}^{D_2}[-,-]$ \\
                          & $\hat{\Sigma}^{B_3}[+,+]$ &
$\hat{\Sigma}^{C_3}[+,-]$ & $\hat{\Sigma}^{D_3}[+,+]$ \\
                          &                           &
$\hat{\Sigma}^{C_4}[-,-]$ & $\hat{\Sigma}^{D_4}[-,+]$ \\
\colrule
$\hat{\Delta}^{A_1}[-,-]$ & $\hat{\Delta}^{B_1}[+,+]$ &
$\hat{\Delta}^{C_1}[+,-]$ & $\hat{\Delta}^{D_1}[-,-]$ \\
$\hat{\Delta}^{A_2}[-,-]$ & $\hat{\Delta}^{B_2}[+,-]$ &
$\hat{\Delta}^{C_2}[-,-]$ & $\hat{\Delta}^{D_2}[+,+]$ \\
                          & $\hat{\Delta}^{B_3}[-,-]$ &
$\hat{\Delta}^{C_3}[-,+]$ & $\hat{\Delta}^{D_3}[-,-]$ \\
                          &                           &
$\hat{\Delta}^{C_4}[+,+]$ & $\hat{\Delta}^{D_4}[+,-]$ \\
\botrule
\end{tabular}}
\label{tab:tab}
\end{table}

\section{Conclusions}
\label{sec:con}

The capability of controlling polarizations of the colliding photons
makes the PLC an ideal machine for probing CP properties of 
neutral spin-zero particles.
To illustrate this capability, we have investigated
the neutral Higgs-boson sector of the MSSM in which
interesting CP-violating mixing could arise via radiative corrections.
First we have considered CP violation in inclusive
$s$-channel Higgs production and constructed three asymmetries 
${\cal A}^{H_i}_{1,2,3}$ for each $H_i$.
%
And we have presented a general formalism for analyzing
CP-violating phenomena in the process $\gamma\gamma \rightarrow \bar{f}f$
by exploiting controllable beam polarizations
and possibly measurable final-fermion polarizations. 
We have shown 
that one can construct more than 20 independent observables for each decay mode
and half of them are CP-odd.


\section*{Acknowledgments}
I wish to thank E. Asakawa, S.Y. Choi, 
K. Hagiwara, J. Ellis, and A. Pilaftsis for valuable
collaborations. 
In particular, I would like to thank S.Y. Choi for careful reading of the
manuscript.
This work was supported in part by
the Korea Research Foundation (KRF)
and the Korean Federation of Science and Technology Societies Grant
and in part by the KRF grant KRF-2005-084-C00001
funded by the Korea Government (MOEHRD, Basic Research Promotion Fund).

\end{document}